\documentclass[twocolumn]{aastex631}
\usepackage{tabularx}

\usepackage{amsmath,amssymb}

\usepackage{soul}

\begin{document}

\shorttitle{Removing pulsars one by one}

\title{
The NANOGrav 15 Yr Data Set: Removing Pulsars One by One from the Pulsar Timing
Array \\
}

\author[0000-0001-5134-3925]{Gabriella Agazie}
\affiliation{Center for Gravitation, Cosmology and Astrophysics, Department of Physics, University of Wisconsin-Milwaukee,\\ P.O. Box 413, Milwaukee, WI 53201, USA}
\author[0000-0002-8935-9882]{Akash Anumarlapudi}
\affiliation{Center for Gravitation, Cosmology and Astrophysics, Department of Physics, University of Wisconsin-Milwaukee,\\ P.O. Box 413, Milwaukee, WI 53201, USA}
\author[0000-0003-0638-3340]{Anne M. Archibald}
\affiliation{School of Mathematics, Statistics, and Physics, Newcastle University, Newcastle upon Tyne, NE1 7RU, UK}
\author{Zaven Arzoumanian}
\affiliation{X-Ray Astrophysics Laboratory, NASA Goddard Space Flight Center, Code 662, Greenbelt, MD 20771, USA}
\author[0000-0002-4972-1525]{Jeremy G. Baier}
\affiliation{Department of Physics, Oregon State University, Corvallis, OR 97331, USA}
\author[0000-0003-2745-753X]{Paul T. Baker}
\affiliation{Department of Physics and Astronomy, Widener University, One University Place, Chester, PA 19013, USA}
\author[0000-0003-0909-5563]{Bence B\'{e}csy}
\affiliation{Department of Physics, Oregon State University, Corvallis, OR 97331, USA}
\author[0000-0002-2183-1087]{Laura Blecha}
\affiliation{Physics Department, University of Florida, Gainesville, FL 32611, USA}
\author[0000-0001-6341-7178]{Adam Brazier}
\affiliation{Cornell Center for Astrophysics and Planetary Science and Department of Astronomy, Cornell University, Ithaca, NY 14853, USA}
\affiliation{Cornell Center for Advanced Computing, Cornell University, Ithaca, NY 14853, USA}
\author[0000-0003-3053-6538]{Paul R. Brook}
\affiliation{Institute for Gravitational Wave Astronomy and School of Physics and Astronomy, University of Birmingham, Edgbaston, Birmingham, B15 2TT, UK}
% ...

\begin{abstract}
Evidence has emerged for a stochastic signal correlated among 67 pulsars within the 15 yr pulsar-timing data set compiled by the NANOGrav collaboration. Similar signals have been found in data from the European, Indian, Parkes, and Chinese pulsar timing arrays. This signal has been interpreted as indicative of the presence of a nanohertz stochastic gravitational-wave background (GWB). To explore the internal consistency of this result, we investigate how the recovered signal strength changes as we remove the pulsars one by one from the data set. We calculate the signal strength using the (noise-marginalized) optimal statistic, a frequentist metric designed to measure the correlated excess power in the residuals of the arrival times of the radio pulses. 
We identify several features emerging from this analysis that were initially unexpected. The significance of these features, however, can only be assessed by comparing the real data to synthetic data sets. After conducting identical analyses on simulated data sets, we do not find anything inconsistent with the presence of a stochastic GWB in the NANOGrav 15 yr data. The methodologies developed here can offer additional tools for application to future, more sensitive data sets. While this analysis provides an internal consistency check of the NANOGrav results, it does not eliminate the necessity for additional investigations that could identify potential systematics or uncover unmodeled physical phenomena in the data.

\end{abstract}

\section{Introduction}
\label{sec:intro}

Pulsar timing arrays (PTAs) can be used to detect gravitational waves (GWs) in the nanohertz frequency band by examining the timing residuals in a network of millisecond pulsars (MSPs) \citep{1978SvA....22...36S, 1979ApJ...234.1100D, 1990ApJ...361..300F}. The timing residuals induced by a Gaussian and isotropic stochastic background of GWs in different pulsars are correlated across the sky in a characteristic way \citep[described by the Hellings-Downs (HD) curve;][]{1983ApJ...265L..39H} which allows them to be distinguished from other mismodelled processes.
It is expected that inspiralling supermassive black holes binaries (SMBHB), with masses in the range $\sim$$£10^8 \,\textrm{--}\, 10^{10}\,M_\odot$ and redshifts $z \lesssim 2$, will produce a stochastic GW background (SGWB) in this frequency band \citep{2003ApJ...583..616J, 2004ApJ...611..623S, 2008MNRAS.390..192S, 2014ApJ...789..156M}, although individual sources and cosmological backgrounds may also be present with comparable amplitudes.

Recently, several PTA collaborations have reported the results of the analyses of their most recent data sets showing the first emerging evidence for a \emph{spatially correlated process} in the data. These results
came from analysing the NANOGrav 15 yr data set \citep[NG15;][]{2023ApJ...951L...8A}, 
the second data release from the European PTA \citep[EPTA2;][]{2023A&A...678A..50E} which includes the first data release from the Indian PTA \citep[][]{2018JApA...39...51J, 2023A&A...678A..50E}, 
the third data release from the Parkes PTA \citep[][]{2023ApJ...951L...6R} and 
the first data release from the Chinese PTA \cite[][]{2023RAA....23g5024X}. 
In the near future, increased sensitivity is expected to be achieved using the forthcoming third data release from the International PTA which will combine many of the aforementioned data sets, data from new observatories such as MeerKAT \citep[which has already made its first data release, MPTA1;][]{2023MNRAS.519.3976M} and eventually the Square Kilometre Array \citep[SKA;][]{2015aska.confE..37J}.
Despite the growing evidence for a stochastic gravitational-wave background (SGWB) there are good reasons to be cautious: tension with previous upper limits \citep[][]{2023ApJ...951L...6R}; hints of unexpected angular correlation in some data sets \citep[EPTA and NG15 monopole;][]{2023A&A...678A..50E, 2023ApJ...951L...8A};
differing signal characteristics between the published data sets and even from different time slices within single data sets \citep[][]{2023A&A...678A..50E, 2023ApJ...951L...6R}.
Given the importance of any detection claim, it is necessary to verify the internal consistency of any data modeling \citep[e.g.,][]{2024arXiv240720510A}, especially for data sets that we intend to combine with others in further analysis. The objective of this paper is to devise tools to carry out such internal consistency checks. This work does not, however, negate the need for additional studies aimed at detecting potential systematics and identifying any unmodeled physical phenomena in the data.

In this paper we investigate the NANOGrav 15 yr data set (NG15 hereafter). This data set is the product of a 67-pulsar PTA (those with a timing baseline $\geq$ 3 years), the most pulsars in any published PTA search currently. In the future, we plan to apply the methods developed here to other data sets, including those under construction within the IPTA.

\section{Executive Summary}
\label{sec:exec_summary}

The NANOGrav collaboration recently published evidence for an SGWB in the 15 yr data set \citep[NG15;][]{2023ApJ...951L...8A}. To deepen our understanding of the findings, we have developed an analysis pipeline that calculates how the recovered signal strength changes as pulsars are cumulatively removed from the PTA. Because our analysis requires the signal strength to be calculated thousands of times, we employ the noise marginalized (NM) optimal statistic (OS), a computationally inexpensive frequentist detection metric, in our assessment of any signal. As well as a signal with the HD spatial correlations expected of an SGWB, we also searched for signals with monopole and dipole spatial correlations. \textit{A priori} we anticipated that the recovered signal strength would exhibit a somewhat gradual decline as pulsars were removed from the array. However, the actual results show abrupt drops and even increases that were not anticipated. In order to quantify how unusual these features were, we designed metrics to quantify them and then compared the data to 100 realizations of simulated PTAs with an SGWB injected and with noise properties based on NG15. 
We find that the results of our pulsar removal analysis of NG15 are consistent with the simulated data sets and, therefore, with an SGWB signal. It is still crucial, however, to pursue further studies that could expose underlying systematics or unmodeled physical influences in the data.

\section{Methods}
\label{sec:methods}

When using PTAs to search for an SGWB, multiple timing noise sources vary over timescales similar to the GW signal (e.g., intrinsic pulsar noise, interstellar-medium–induced radio-frequency-dependent fluctuations, and timing-model errors). 
To disentangle signal from noise, a joint estimate of both can be calculated using a Bayesian framework. 
By performing a fully Bayesian GW search, the analysis considers the relationships between parameters, provides a more nuanced understanding of the underlying astrophysical processes and incorporates all sources of uncertainty.
However, a fully Bayesian analysis that incorporates HD spatial correlations requires substantial computing resources. 
Therefore, a useful complement to the Bayesian framework is the computationally inexpensive frequentist detection metric known as the OS \citep[][]{PhysRevD.79.084030}.
\subsection{The Optimal Statistic}
\label{subsec:os}

The vector of timing residuals in each pulsar is denoted $\delta \mathbf{t}_a$, where the index $a$ labels the different pulsars in the array.
When looking for a stochastic signal process that is spatially correlated between different pulsars, the cross-covariance between two pulsars ($a\neq b$) is used;
\begin{align} \label{eq:residuals}
	\rho_{ab} = \left< \delta \mathbf{t}_a \delta \mathbf{t}_b \right>.
\end{align}

The OS is a frequentist estimator for the amplitude of the correlated (i.e.,\ signal) process.
The optimal amplitude statistic, $\hat{A}$, is defined as the value of the amplitude parameter $A$ that minimizes the quantity
\begin{align}\label{eq:OS}
	\chi^2 = \sum_{a, b<a} \left(\frac{\rho_{ab}-A^2\Gamma_{ab}}{\sigma_{ab}}\right)^2\,,
\end{align}
where $\sigma_{ab}$ are the uncertainties in the cross-correlations and where $\Gamma_{ab}$ is the overlap reduction function (ORF; which quantifies the degree of correlation between pulsars as a function of their positions on the sky.)
Under the assumptions that the noise is Gaussian and is much greater than the signal it can be shown that this statistic is optimal in the sense that it maximizes the signal-to-noise ratio (S/N). Note that the autocorrelation terms involving $\rho_{aa}$ are intentionally omitted from the summation in this definition.

The OS defined above assumes a known ORF for a specific spatial correlation.
However, in addition to the HD spatial correlations induced by an SGWB, other sources of spatially correlated timing residuals are possible. 
For example, clock errors can lead to monopole spatial correlations \citep[]{2012MNRAS.427.2780H} and uncertainty in the position of the solar system barycenter can produce dipole spatial correlations \citep[]{2010ApJ...720L.201C} between the pulsars. 
The OS can be generalized to allow for a mixture of different possible spatial correlations \citep{2023PhRvD.108l4081S}.
These multiple-component optimal statistic (MCOS) amplitudes, $A^\alpha$, are now the values that minimize
\begin{align}\label{eq:MCOS}
	\chi^2 = \sum_{a, b<a} \left(\frac{\rho_{ab}-\sum_\alpha A_\alpha^2\Gamma^\alpha_{ab}}{\sigma_{ab}}\right)^2,
\end{align}
where $\Gamma^\alpha$ is the ORF for each component. \citet{2023PhRvD.108l4081S} show that using the MCOS results in a more precise recovery of spatially correlated signals compared to the OS, and eliminates the problem of overestimating the amplitude of correlations not present in the data. Three different spatial correlations are considered here: a monopole correlation, a dipole correlation and the HD correlation which is predominantly quadrupolar, i.e.,\ $\alpha\in\{\mathrm{Monopole},\mathrm{Dipole},\mathrm{HD}\}$.
The ORFs for these spatial correlations are respectively
\begin{align} 
	&\Gamma^{\rm Monopole}_{ab} = 1, \label{eq:monopoleORF} \\ 
    &\Gamma^{\rm Dipole}_{ab} = \cos\xi_{ab},\quad \mathrm{and} \label{eq:dipoleORF} \\ 
	&\Gamma^{\rm HD}_{ab} = \frac{\delta_{ab}}{2}+\frac{1}{2}-\frac{(1-\cos\xi_{ab})}{4}\bigg(\frac{1}{2}\nonumber \\&\hspace{1cm}-3\log\left[\frac{1-\cos\xi_{ab}}{2}\right]\bigg) \label{eq:HDORF},
\end{align}
where $\xi_{ab}$ is the angle on the sky between pulsar $a$ and pulsar $b$. 

In this paper, both the single component (SC) OS (see Eq.~\ref{eq:OS}) and MCOS (see Eq.~\ref{eq:MCOS}) are used. SCOS analyses are performed one at a time using each of the three ORFs in Eqs.~\ref{eq:monopoleORF} to \ref{eq:HDORF}.  Whenever the MCOS is used, it includes all three ORFs simultaneously.

The OS and the MCOS are computationally cheap to evaluate (for a typical PTA data set they can be evaluated in $\approx 0.5 \; \mathrm{s})$. This characteristic makes them ideal for iterative analyses and facilitates the comprehensive exploration of a range of scenarios, such as different combinations of pulsars and spatial correlations and also when marginalizing over pulsar-noise parameters (see Section \ref{subsec:nmos}).

\subsection{The NMOS}
\label{subsec:nmos}
In cases where pulsars exhibit significant red noise, the OS yields biased results due to the strong covariance between individual red-noise parameters and the amplitude of the SGWB. This bias can be mitigated by marginalizing over the individual pulsars’ red-noise parameters using the posterior probability distributions from a Bayesian analysis of all PTA pulsars. This Bayesian step includes an SGWB red-noise model that has a spectrum common to all pulsars, but that is agnostic regarding any spatial correlations between pulsars. This step must not be confused with a Bayesian analysis complete with spatial correlations, which is a much more computationally expensive process \citep[]{2018PhRvD..98d4003V}. In practice, each time the OS is calculated, the pulsars' individual red-noise parameters are drawn from the posterior distributions of the spatially uncorrelated Bayesian analysis; we calculate the OS 10,000 times, resulting in a distribution from which an average OS can be calculated. This method more accurately determines the strength of a signal embedded in the pulsar timing residuals than a single OS calculation and is many orders of magnitude faster than performing a Bayesian analysis complete with spatial correlations. We, therefore, use the NMOS throughout this work. 

\subsubsection{Spatially Uncorrelated Bayesian Analysis}
\label{sec:bayesian}

As described in Section~\ref{subsec:nmos}, to calculate the NMOS we require the posterior probability distributions that result from a spatially uncorrelated Bayesian analysis of the pulsar data, which ignores spatial correlations, and we briefly describe that analysis here. When looking for the effects of gravitational waves, the pulsar data of interest are the timing residuals, which represent the deviations between observed and predicted pulse arrival times. For each pulsar we use the NG15 pulse times of arrival (TOAs) and the NG15 noise models
to compute the timing residuals. We assume that the timing residuals of the pulsars are produced by various physical processes characterized by variable parameters. Bayes' theorem is then employed to deduce the probability distribution of these parameters of interest. Under the assumption of accurate noise modeling for each pulsar, the physical processes assumed to give rise to systematic trends seen in a pulsar's timing residuals are:
\begin{enumerate}
    \item{The SGWB. The SGWB will imprint a common spectrum on all pulsar timing residuals. We model this SGWB spectrum as a $-2/3$ characteristic-strain power law, translating to a $-13/3$ power law for timing residuals. This power law is expected if the background is due to the incoherent superposition of radiation from SMBHBs in circular orbit in-spiraling due to loss of energy and angular momentum produced by GW emission ~\citep{2001astro.ph..8028P}. The amplitude of the SGWB spectrum is left as a free parameter and we assume a log-uniform prior probability distribution between $10^{-18}$ and $10^{-11}$.}
    \item{Individual pulsar red noise (also known as timing noise). Suggested interpretations of pulsar red noise include intrinsic changes in the spindown rate of the pulsar \citep[]{1994ApJ...428..713K}, interstellar propagation effects \citep[]{1984Natur.307..527A, 1990ARA&A..28..561R, 1994ApJ...428..713K, 1995A&A...296..169C}, and the presence of a circumpulsar asteroid belt \citep[]{2013ApJ...766....5S}. The red noise for each pulsar is also modeled as a power law. The amplitude and the slope of the spectrum are both free parameters. We assume a log-uniform prior probability distribution between $10^{-20}$ and $10^{-11}$ for the amplitude of the red noise and a uniform prior probability distribution between 0 and 7 for its slope.}
\end{enumerate}
We use Markov chain Monte Carlo techniques to sample randomly from the joint posterior distribution of our model parameters. Runs were performed using the \texttt{PTMCMC} sampler \citep{2017zndo...1037579E}.

Each time a pulsar is removed from the PTA (as described in the next section) this Bayesian analysis step is completed anew with fewer pulsars. The red-noise parameters of each pulsar subarray are then drawn from the Bayesian posterior distributions as described in Section \ref{subsec:nmos}.

\subsection{Pulsar removal analysis}
\label{subsec:subsets}

Our aim is to use the above methods to quickly search for a correlated signal in the timing residuals of a PTA. Any signal detected will vary as a function of the number of pulsars in the array. One by one we eliminate pulsars from the full array and observe how the recovered signal is affected. The order in which we removed pulsars is based on their noise levels, specifically the weighted root-mean-square of epoch-averaged post-fit timing residuals after whitening (i.e., any red-noise contribution is removed) as listed in Table 6 of \citet[][]{2023ApJ...951L...9A}\footnote{For PSR~B1937+21, the
value listed in Table 6 of \citet[][]{2023ApJ...951L...9A}
is incorrect.
This was discovered in the latter stages of this analysis.
As a result, PSR~B1937+21 appears at a position in the pulsar removal ordering that reflects the value quoted in the \citet[][]{2023ApJ...951L...9A} table.} We implemented two distinct exclusion methodologies for the NG15 pulsars: one where the least noisy pulsars were excluded first, and another where the most noisy pulsars were excluded first. \textit{A priori} we anticipate that the removal of the least noisy pulsars will have the most pronounced impact on the signals recovered from the timing residuals. It is for this reason that we prioritize the scenario in which the least noisy pulsars are removed first. The case where we remove the most noisy pulsars first is discussed in Appendix \ref{app:noisy_first} of this paper.

\section{Results}
\label{sec:results}

\subsection{The Full Array}
\label{subsec:all_pulsars}

\begin{figure*}
    \centering
    \includegraphics[width=0.95\textwidth]{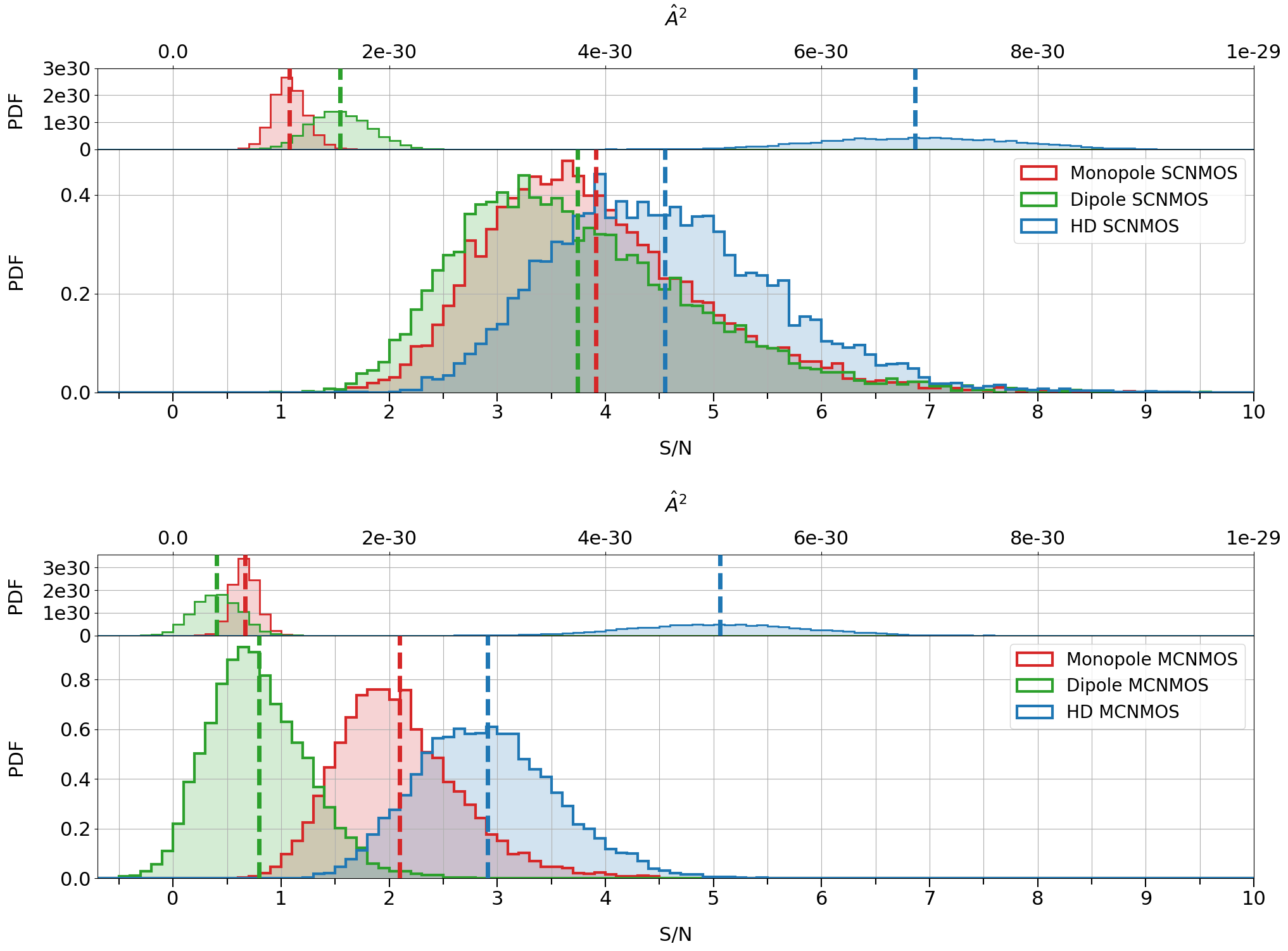}
    \caption{\label{fig:full_array_stats}
        \emph{Top plot:} results for the SCNMOS.
        \emph{Bottom plot:} results for the MCNMOS.
        In both cases the narrow top panels shows the distribution of the squared amplitudes $\hat{A}^2$ for the monopole, dipole and HD components, while the wider bottom panels show the distribution of the S/N for the same three spatial correlations. The values were calculated 10,000 times as part of the noise marginalization process. Dashed vertical lines indicate the mean value for a distribution. The mean (and not the median) is used here to allow direct comparison with values calculated in NG15.
    }
\end{figure*}

We begin by calculating the signal found in the timing residuals of the full NG15 timing array comprised of 67 pulsars. 
The results from both the single-component SCNMOS and the MCNMOS analyses are shown in Table~\ref{tab:allpulsars} and Figure~\ref{fig:full_array_stats}, and were calculated for monopole, dipole and HD spatial correlations. Both our SCNMOS and MCNMOS results for the full array are consistent with those of NG15 (Figure 4 and Table 2 in \citet[][]{2023ApJ...951L...8A} respectively). As we are using the full array in this case, this also permits us to compare our results with the fully Bayesian analysis that includes HD spatial correlations in the \citet[][]{2023ApJ...951L...8A} paper. Again they are in agreement.

\begin{table*}
    \begin{tabular}{|c|c|c|c|c|c|c|}
        \cline{2-7}
        \multicolumn{1}{c|}{} & \multicolumn{2}{c|}{Monopole Correlations} & \multicolumn{2}{c|}{Dipole Correlations} & \multicolumn{2}{c|}{HD Correlations} \\
        \cline{2-7}
        \multicolumn{1}{c|}{} & $\hat{A^{2}}$ & S/N & $\hat{A^{2}}$
        & S/N & $\hat{A^{2}}$ & S/N \\
        \hline
        SCNMOS & 1.08 $\times 10^{-30}$ & 3.91 & 1.55 $\times 10^{-30}$ & 3.74 & 6.87 $\times 10^{-30}$ & 4.55 \\
        \hline
        MCNMOS & 6.65 $\times 10^{-31}$ & 2.09 & 4.05 $\times 10^{-31}$ & 0.79 & 5.06 $\times 10^{-30}$ & 2.91 \\
        \hline
    \end{tabular}
    \caption{Mean values for the OS $\hat{A}$ and S/N for the three ORFs and for both SCNMOS and MCNMOS, taken from Figure~\ref{fig:full_array_stats}. These are calculated when all 67 NANOGrav pulsars are included in the PTA. The mean (and not the median) is used here to allow direct comparison with the values calculated in NG15.
    }
    \label{tab:allpulsars}
\end{table*}

\subsection{Removing pulsars: SC analysis}
\label{subsec:SC}
\begin{figure*}
    \centering
    \includegraphics[width=\textwidth]{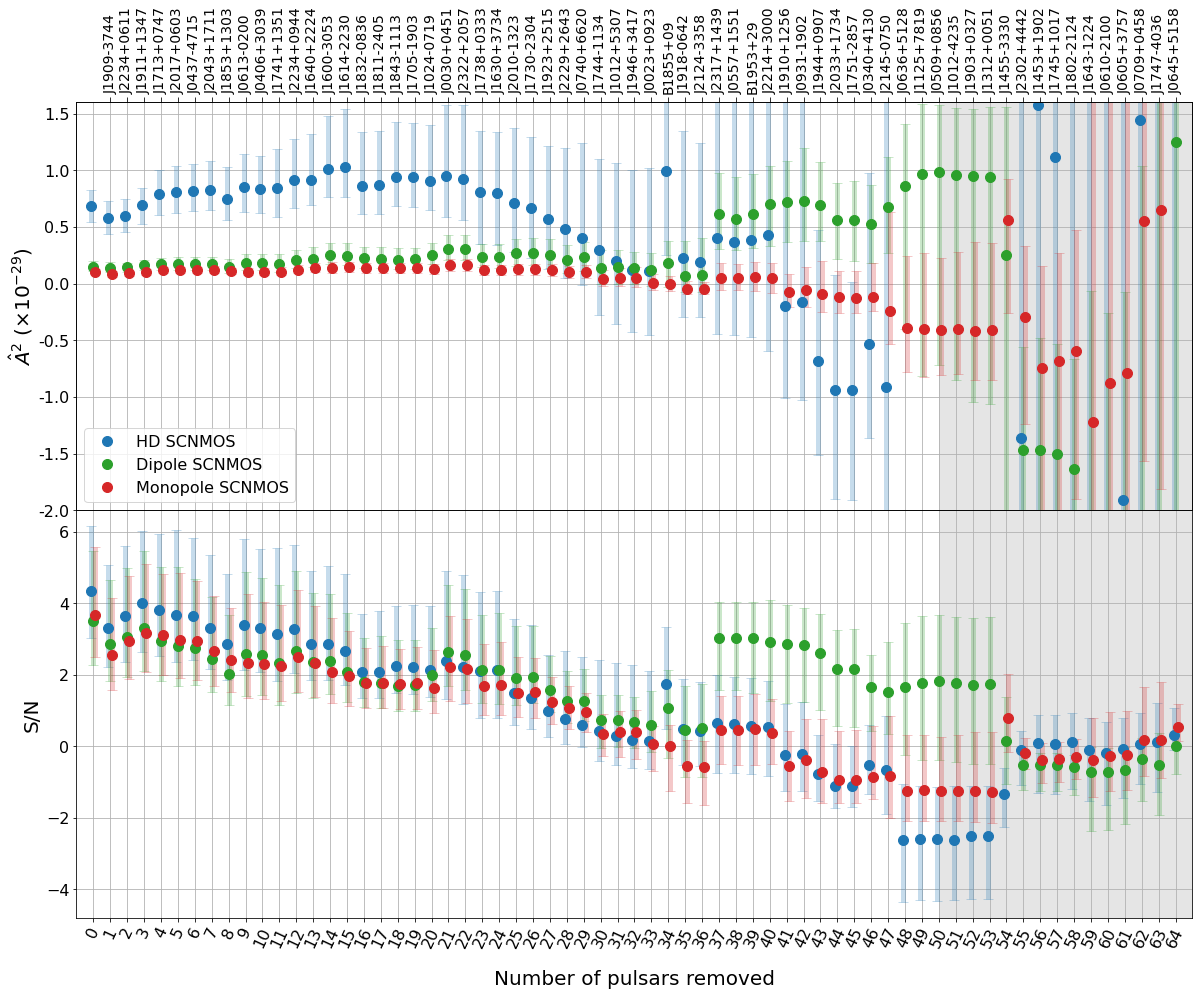}
    \caption{\label{fig:NG15_SC_bestfirst}
        The evolution of the SCNMOS squared amplitude $\hat{A}^2$ and its S/N for NG15 as the least noisy pulsars are removed one by one from the PTA.
        The bottom axis is the number of pulsars removed
        while the top axis gives the name of most recent pulsar to have been removed.
        Each circle in the upper panel represents the median of the 10,000 SCNMOS $\hat{A}^2$ distribution calculated for a specific subset of the entire PTA. The bars attached to the circles represent the range between the 5th and 95th percentiles of the $\hat{A}^2$ distribution. In the lower panel the circles and bars depict the median and range respectively of the S/N values of the SCNMOS analysis. Note that one cannot assess the significance of the changes in the median circles by referring to the bars depicting the spread of the distribution from which it was calculated; the uncertainty on the median is of order 100 ($\sqrt{10,000}$) times smaller than the width of the range bars. The data within the gray-shaded region are not considered when calculating the metrics defined in Section~\ref{subsec:metrics} as the behavior of $\hat{A}^2$ and S/N can become increasingly erratic due the small number of low-sensitivity pulsars remaining.
    }
\end{figure*}

We next remove the least noisy pulsars from the full PTA one by one. As seen in the upper panel of Figure~\ref{fig:NG15_SC_bestfirst} the SCNMOS for the dipole and monopole spatial correlation is around $10^{-30}$ and remains relatively stable until approximately half of the pulsars have been excised from the full PTA. Upon the removal of the 37th pulsar (PSR~J2317$+$1439) the dipole spatial correlation OS suddenly increases. The SCNMOS for the HD spatial correlation remains relatively flat until approximately a third of the pulsars have been removed. At this point it experiences a steady decline until an abrupt increase upon the removal of the 34th pulsar (PSR~B1855$+$09), followed by an equally abrupt drop as the next pulsar is removed. In the lower panel of Figure~\ref{fig:NG15_SC_bestfirst} all three spatial correlations begin with an S/N of around 4 and then undergo a relatively steady drop to an S/N around zero after the removal of 33 pulsars. However, there is a significant drop in S/N (especially for HD and monopole spatial correlations) when the first pulsar is removed. The jumps seen in the OS after the removal of pulsar 34 and 37 are reflected in the S/N for the HD and dipole spatial correlations respectively.

\subsection{Removing pulsars: MC analysis}
\label{subsec:MC}

\begin{figure*}
    \centering
    \includegraphics[width=\textwidth]{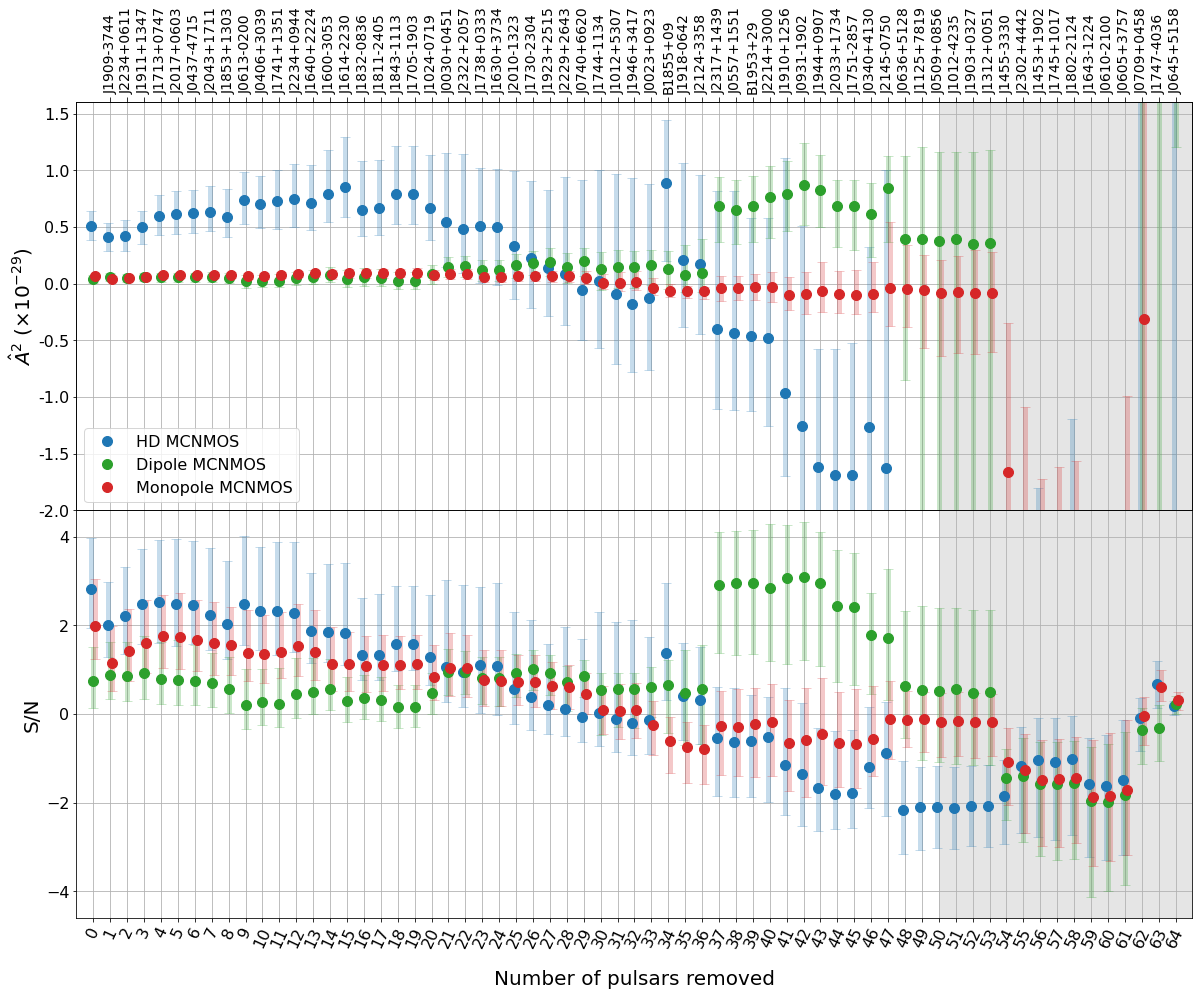}
    \caption{\label{fig:NG15_MC_bestfirst}
        The evolution of the MCNMOS and S/N for NG15 as the least noisy pulsars are removed one by one from the PTA.
        Similar to Figure~\ref{fig:NG15_SC_bestfirst}; see that caption for more details.
    }
\end{figure*}

As expected, the main features described above and seen in Figure~\ref{fig:NG15_SC_bestfirst} also appear in Figure~\ref{fig:NG15_MC_bestfirst}. The S/N values, however, are systematically lower for the MC analysis compared to the SC analysis.

As the MCOS more accurately recovers spatially correlated signals than the OS \citep[][]{2023PhRvD.108l4081S}, we have identified four features of interest in the MCNMOS pulsar removal plots of Figure~\ref{fig:NG15_MC_bestfirst}. The features are described and quantified in the following and are highlighted in Figure~\ref{fig:all_simulations}. Although we focused on these four features, it is important to note that various others could have been selected; we primarily chose these particular features based on notable characteristics observed in the plotted data.

\emph{Feature (I)}. Initial S/N for the HD spatial correlations.
The median S/N of the signal with HD spatial correlations when all 67 NANOGrav pulsars are included in the pulsar timing array is $\sim$2.8. We see a value close to this in Table 2 of \cite{2023ApJ...951L...8A}.

\emph{Feature (II)}. S/N drop after removal of one pulsar for the HD spatial correlations.
When we remove the least noisy pulsars first, an S/N drop from $\sim$2.8 to $\sim$2.0 is seen in the HD spatial correlations plot when the first pulsar PSR~J1909$-$3744 is removed. This drop of 29\% after the removal of only one pulsar from the PTA appears notably large. In the highly idealized case where all pulsars are identical, the S/N is expected to scale linearly with the number of pulsars in the array ~\citep[see e.g., Equation 24 from][]{2013CQGra..30v4015S}; a 29\% drop is clearly much larger than this naive approximation would predict. However, PSR~J1909$-$3744 is the most precisely timed pulsar and so its removal would be expected to produce a larger drop in S/N compared to other pulsars. Specifically, the weighted root-mean-square of epoch-averaged post-fit timing residuals after whitening for PSR~J1909$-$3744 is 0.066 $\mu$s \citep{2023ApJ...951L...9A}, which is $\sim$43 times smaller than the highest value in NG15. Nevertheless, the magnitude of the drop still seems surprising for a single pulsar removal and its significance will be quantified later in this paper.

\emph{Feature (III)}. Sharp Increase in S/N for HD spatial correlations.
When 33 pulsars have been removed in the order of increasing noise, the S/N of the HD spatial correlation is $\sim$-0.1. On the removal of the 34th pulsar (PSR~B1855$+$09) the S/N for the PTA consisting of the remaining 33 pulsars rises to $\sim$1.4. \textit{A priori} the S/N would be expected to decrease when a pulsar is removed. It is surprising, therefore, that the removal of a single pulsar could produce such a significant S/N increase.

\emph{Feature (IV)}. Sharp Increase in S/N for dipole spatial correlations.
The S/N of the dipole spatial correlation jumps from $\sim$1 to $\sim$3 when the 37th pulsar (PSR~J2317+1439) is removed in the order of increasing noise.\\

In order to evaluate the significance of these four features and to determine whether they indicate some unmodelled aspect of the NG15 data set, we repeat the MCNMOS calculations described in this paper on 100 simulated data sets with properties statistically akin to those of NG15. The analysis of the real and the simulated data sets are compared in the following section.

\section{Comparison with simulated data}
\label{subsec:sim_data}

To assess the peculiarity of the prominent features we see in our MCNMOS analysis of NG15, we make a comparison with an ensemble of simulated data sets that share the statistical properties of NG15. 
The simulated data were generated in \citet[][]{2023ApJ...951L...8A} following \citet{2021ApJ...911L..34P}. 
We note that the construction of these simulated data sets is central to the conclusions of this paper. Comprehensive details of their construction can be found in these previous references, but to summarize, each of the 100
simulated data sets share the same pulsars, timing cadence and noise parameters as NG15 and also include a spatially correlated power-law SGWB with power-law amplitude $2.7\times10^{-15}$ and spectral index $-13/3$. The data sets differ only in their random-noise realizations. 
For each realization, we calculated the MCNMOS and removed pulsars one by one as in the analysis of NG15. By performing the same analysis on simulated data sets we are able to determine the expected behavior of the MCNMOS as pulsars are steadily removed. By comparing the results from the NG15 data with those from the ensemble of simulated data it is possible to assign a significance to the unexpected features identified in Section~\ref{subsec:MC}. Figure~\ref{fig:sim_MC_bestfirst} shows the results for a single representative simulated data set with the least noisy pulsars removed first; this can be compared to the equivalent plot from the NG15 data set, shown in Figure~\ref{fig:NG15_MC_bestfirst}.

\subsection{Defining metrics}
\label{subsec:metrics}
\begin{figure*}
    \centering
    \includegraphics[width=0.95\textwidth]{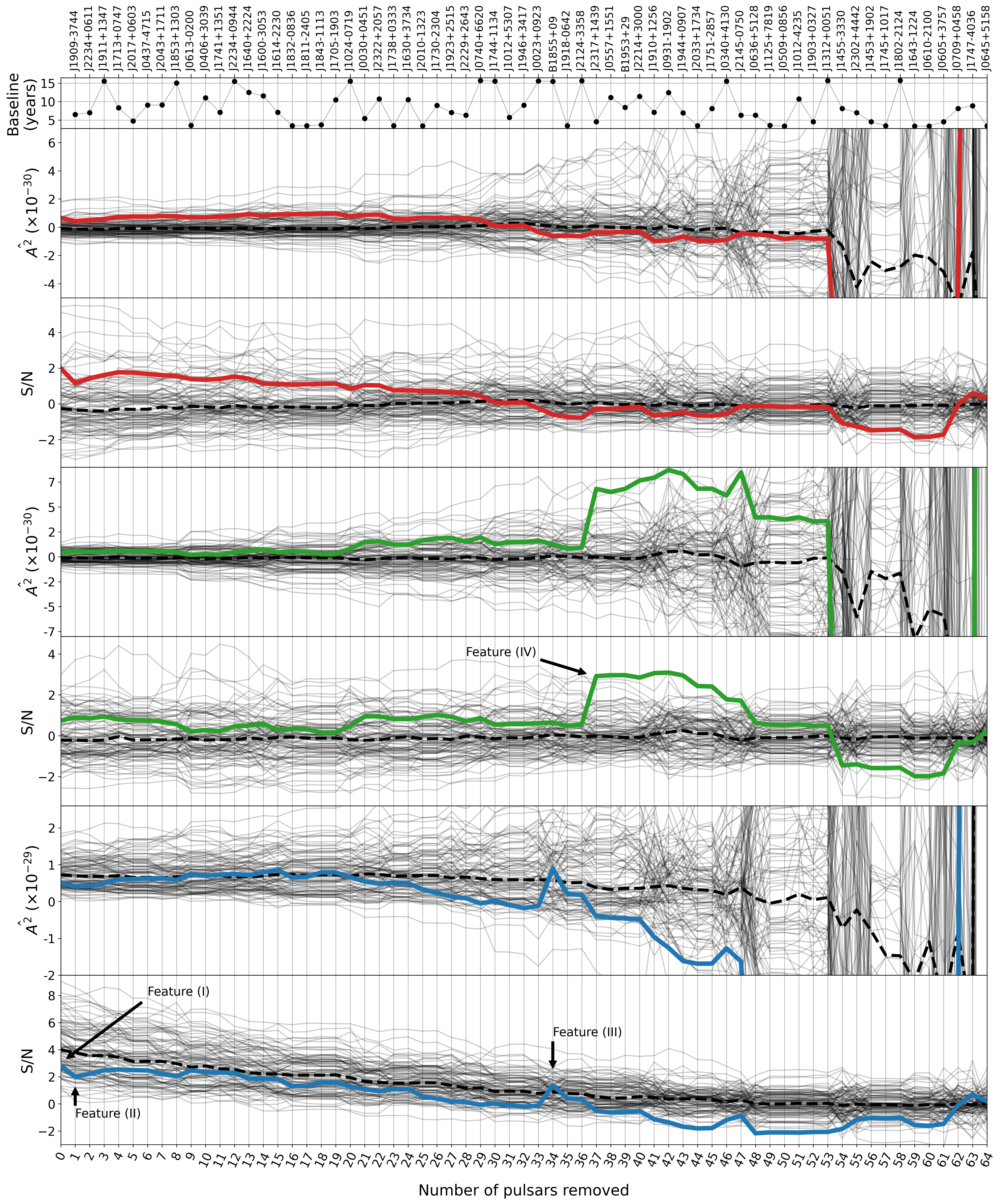}
    \caption{\label{fig:all_simulations}
        The top panel shows the length of the timing baseline for each NG15 pulsar. The lower six panels are in pairs for each spatial correlation, tracking the MCNMOS and the S/N as the least noisy pulsars are removed one by one from the PTA. The panels with a red line correspond to monopole spatial correlations. The panels with a green line correspond to dipole spatial correlations. The panels with a blue line correspond to HD spatial correlations. The colored lines are the results from NG15 and each gray trace corresponds to one of the 100 simulated data sets. The black dashed lines show the medians of the 100 gray traces. The features described in Section~\ref{subsec:MC} are highlighted.
    }
\end{figure*}

\begin{figure*}
    \centering
    \includegraphics[width=0.95\textwidth]{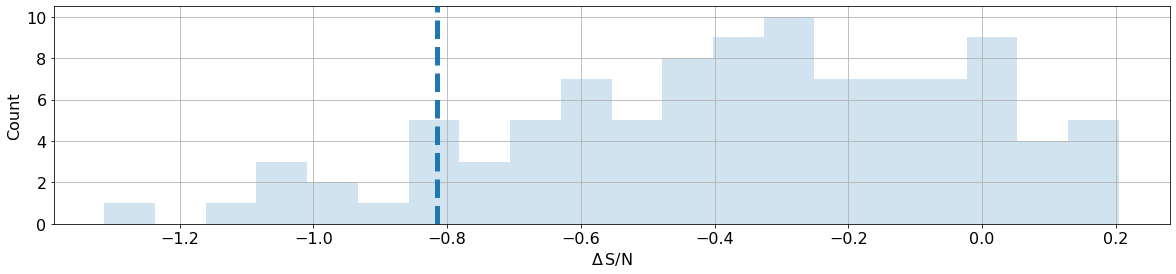}
    \caption{\label{fig:paper_drop_hd_sn}
The distribution of the drop in S/N when the first pulsar is removed, for signals with HD spatial correlations. The histogram shows the median values from the MCNMOS analysis for each of the 100 simulated data sets. The dashed vertical line indicates the median value from the MCNMOS analysis of NG15.
    }
\end{figure*}

To quantify the features discussed in Section~\ref{subsec:MC} we define three metrics:
\begin{enumerate}
  \item The S/N with all 67 pulsars in the array.
  \item The decrease in S/N after the first pulsar (PSR~J1909$-$3744) is removed.
  \item The largest increase in S/N produced by the removal of any one pulsar.
\end{enumerate}
The metrics focus on S/N because each of our four features in Section~\ref{subsec:MC} appear in S/N analyses. For completeness, however, we compute these metrics for $\hat{A}$ as well as the S/N. For each of these two variables, we calculate the three metrics for each of the three ORFs, resulting in a total of 18 metric values.
These 18 values are calculated for NG15 and for the 100 simulations (see Figure~\ref{fig:paper_drop_hd_sn}, which shows the example of
the distribution of the drop in S/N when the first pulsar is
removed, for signals with HD spatial correlations). This allows us to compute $p$-values thereby determining the probability that a given NG15 pulsar removal feature would appear, given the null hypothesis that NG15 is accurately modeled and internally consistent. The 18 $p$-values are shown in Table~\ref{tab:p_values}; three of the 18 are less than or equal to 0.01. When calculating the metrics we only consider the data until the point where 50 pulsars have been removed; this is indicated in Figures \ref{fig:NG15_SC_bestfirst}, \ref{fig:NG15_MC_bestfirst} and \ref{fig:sim_MC_bestfirst} by the gray regions. This cutoff is somewhat arbitrary, but is justified by the fact that after this point, the behavior of $\hat{A}^2$ and S/N becomes increasingly erratic because of the small number of low-sensitivity pulsars remaining. This is judged by eye from e.g., Figures~\ref{fig:NG15_SC_bestfirst}, \ref{fig:NG15_MC_bestfirst} and \ref{fig:all_simulations}.

\begin{table*}
  \begin{tabular}{|c|c|c|c|}
    \hline
     Variable & Value with all pulsars in array & Drop on first removal & Largest positive jump \\
    \hline
    \hline
    Monopole $\hat{A}$ & 0.92 & $< 0.01$ & 0.96 \\
    \hline
    Monopole S/N & 0.90 & $< 0.01$ & 0.87 \\
    \hline
    Dipole $\hat{A}$ & 0.75 & 0.78 & 0.14 \\
    \hline
    Dipole S/N & 0.74 & 0.67 & 0.01 \\
    \hline
    HD $\hat{A}$ & 0.32 & 0.14 & 0.59 \\
    \hline
    HD S/N & 0.25 & 0.10 & 0.07  \\
    \hline
  \end{tabular}
  \caption{The $p$-values for the 18 metrics, calculated by comparing the NG15 data to the 100 simulated data sets.}
  \label{tab:p_values}
\end{table*}

\section{Discussion}
\label{sec:discussion}

The behavior of the MCNMOS as applied to NG15, shown in Figure~\ref{fig:NG15_MC_bestfirst} and described in Section~\ref{sec:results}, exhibits several initially unexpected features as pulsars are steadily removed from the array. To contextualize the features and to determine whether they point to some unmodeled processes in the NG15, we have repeated the analysis using simulated data with known embedded GW signal and noise properties. To quantify the significance of the unexpected NG15 features we calculate $p$-values for 18 applications of the three metrics described in Section~\ref{subsec:metrics}. Out of 18 instances, there are three with $p$-values less than or equal to 0.01.

We conducted the removal of pulsars from the array in two distinct sequences, acknowledging that a vast number of other permutations could be explored.

We examined the timing baseline duration for each pulsar (refer to Figure~\ref{fig:all_simulations}) to investigate any potential correlation with $\hat{A}$ and S/N as pulsars were sequentially removed from the timing array. Our analysis revealed no apparent correlation.
However, removing pulsars can decrease the time spanned by a PTA's constituent observations. This consequently affects the frequency range over which the PTA can detect a signal. In this analysis, we maintain a constant number of frequency components in the gravitational wave background (GWB) model. This decision is justified by the fact that, as pulsars are removed in the specific sequence chosen for this work, the observation span decreases by less than 60 days. This reduction is minimal compared to the total timespan of 16 years for the complete 67-pulsar PTA.

\emph{Dropout analysis} \citep{2019ApJ...880..116A, 2020ApJ...905L..34A} was employed in NG15 to evaluate the significance of the GWB signal in individual pulsars. The dropout method introduces a binary parameter to either include or exclude the common signal for a specific pulsar. The dropout factor or the Bayes factor then indicates the degree to which each pulsar supports the presence of the GWB signal. In our approach, instead of evaluating each pulsar independently we removed pulsars in a cumulative manner. While both methods aim to understand the robustness and consistency of the detected signal, the dropout method focuses on the individual contribution of each pulsar in isolation. In contrast, our cumulative removal approach provides insight into the collective impact on the signal as pulsars are removed and so no single removal (other than the first) can be viewed in isolation. Comparing Figures~\ref{fig:NG15_SC_bestfirst} with \ref{fig:NG15_SC_noisyfirst} and \ref{fig:NG15_MC_bestfirst} with \ref{fig:NG15_MC_noisyfirst} we see that the effect of removing any particular pulsar can be quite different depending on the order of removal. Furthermore, Figure~\ref{fig:all_simulations} demonstrates that even when the order of removal is preserved, the removal of a particular pulsar can have a different effect on the recovered signal for different realizations of the simulation.

A recently identified limitation of the OS and its derivatives is the introduction of a bias when a correlated signal is significant. The bias arises from the overlooked correlations between pulsar pairs 
\citep[][see also the discussion in Appendix C of \citealp{2023PhRvD.108l4081S}]{2023PhRvD.108d3026A}.
As we have not considered these correlations, the reported values for $\hat{A}^2$ will be affected by the bias, which will decrease as more pulsars are removed and the signal strength drops. We use the OS in this analysis because it is a well established statistic which was widely used in NG15, allowing us to make direct comparisons. Additionally, the techniques we employed are computationally inexpensive and so enabled us to calculate the recovered signal strength thousands of times within a reasonable time frame. While we recognize the limitations of the OS, we mitigate them by comparing the results obtained from real data with those from simulated  data sets. This comparative analysis provides important context and allows meaningful assessments.

Another recent expansion of the OS is the Per-Frequency Optimal Statistic \citep{2024arXiv240611954G} that enables frequency-by-frequency estimation of the GWB spectrum.

We acknowledge that we have selected to investigate only certain outstanding features that result from our analysis, specifically those which seem to contradict the null hypothesis that the NG15 analysis is comprehensively modeled and internally consistent. Such selective analysis will have a significant impact on the calculation of our $p$-values, making them smaller than they would otherwise be. Nevertheless it is worthwhile to notice these features, quantify how often we expect them to occur and keep them in mind when conducting further analyses.

\section{Conclusions and Future Work}
\label{sec:conclusions}

We have investigated the behavior of recovered signal strength when pulsars are removed one by one from NG15. Despite variations during pulsar removal that we had not anticipated, these fluctuations are consistent with those observed in simulated data sets that have matching noise properties and a GWB injected. After calculating $p$-values for the data, we find that the results are consistent with the simulated data sets and, therefore, with an SGWB signal. The purpose of our analysis was to provide internal consistency checks on the NANOGrav 15 yr data set, and the results indicate that the data have successfully passed these tests. It also remains essential to conduct further investigations that may reveal systematics or unmodeled physical effects within the data.

The tools developed here have proved to be useful for checking the internal consistency of PTA data. Although we have applied them to NANOGrav 15 yr data set in this work, these tools could be applied routinely to new PTA data sets as they become available. We plan to analyze the EPTA DR2 data set next, with an eye on IPTA DR3 and MeerKAT data.

\section*{Author contributions}
An alphabetical order author list was used for this paper in
recognition of the fact that a large, decade-timescale project
such as NANOGrav is necessarily the result of the work of
many people. All authors contributed to the activities of the
NANOGrav Collaboration leading to the work presented here
and reviewed the manuscript, text, and figures prior to the
paper’s submission. Additional specific contributions to this
paper are as follows. P.R.B., H.M., C.J.M., and A.V. collaboratively developed the concept and main ideas for this work. P.R.B. was responsible for data analysis and creating the figures. P.R.B., J.S.H., H.M., C.J.M., S.R.T., A.V., S.J.V., and C.A.W. contributed to writing the manuscript. N.S.P. created the simulated PTA  data sets. G.A., A.A., A.M.A., Z.A., P.T.B., P.R.B.,
H.T.C., K.C., M.E.D., P.B.D., T.D., E.C.F., W.F., E.F., G.E.F.,
N.G., P.A.G., J.G., D.C.G., J.S.H., R.J.J., M.L.J., D.L.K., M.
K., M.T.L., D.R.L., J.L., R.S.L., A.M., M.A.M., N.M., B.W.M., C.N., D.J.N., T.T.P., B.B.P.P., N.S.P., H.A.R., S.M.R., P.S.R., A.S., C.S., B.J.S., I.H.S., K.S., A.S., J.K.S., and H.M.W. developed the 15 yr data set through a combination of
observations, arrival time calculations, data checks and
refinements, and timing model development and analysis;
additional specific contributions to the data set are summarized
in \citet{2023ApJ...951L...9A}. J.S.H., J.G., N.L., N.S.P., J.P.S., and J.K.S. performed noise analyses on the data set.

\section*{Acknowledgments}

L.B. acknowledges support from the National Science Foundation under award AST-1909933 and from the Research Corporation for Science Advancement under Cottrell Scholar Award No. 27553.
P.R.B. is supported by the UK's Science and Technology Facilities Council, grant number ST/W000946/1.
S.B. gratefully acknowledges the support of a Sloan Fellowship, and the support of NSF under award \#1815664.
M.C. and S.R.T. acknowledge support from NSF AST-2007993.
M.C. and N.S.P. were supported by the Vanderbilt Initiative in Data Intensive Astrophysics (VIDA) Fellowship.
Support for this work was provided by the NSF through the Grote Reber Fellowship Program administered by Associated Universities, Inc./National Radio Astronomy Observatory.
K.C. is supported by a UBC Four Year Fellowship (6456).
M.E.D. acknowledges support from the Naval Research Laboratory by NASA under contract S-15633Y.
T.D. and M.T.L. are supported by an NSF Astronomy and Astrophysics Grant (AAG) award number 2009468.
E.C.F. is supported by NASA under award number 80GSFC21M0002.
G.E.F., S.C.S., and S.J.V. are supported by NSF award PHY-2011772.
K.A.G. and S.R.T. acknowledge support from an NSF CAREER award \#2146016.
A.D.J. and M.V. acknowledge support from the Caltech and Jet Propulsion Laboratory President's and Director's Research and Development Fund.
A.D.J. acknowledges support from the Sloan Foundation.
The work of N.La., X.S., and D.W. is partly supported by the George and Hannah Bolinger Memorial Fund in the College of Science at Oregon State University.
N.La. acknowledges the support from Larry W. Martin and Joyce B. O'Neill Endowed Fellowship in the College of Science at Oregon State University.
Part of this research was carried out at the Jet Propulsion Laboratory, California Institute of Technology, under a contract with the National Aeronautics and Space Administration (80NM0018D0004).
D.R.L. and M.A.M. are supported by NSF \#1458952.
M.A.M. is supported by NSF \#2009425.
H.M. is supported by the UK Space Agency, Grant No. ST/V002813/1, ST/X002071/1 and ST/Y004922/1.
C.M.F.M. was supported in part by the National Science Foundation under Grants No. NSF PHY-1748958 and AST-2106552.
A.Mi. is supported by the Deutsche Forschungsgemeinschaft under Germany's Excellence Strategy - EXC 2121 Quantum Universe - 390833306.
The Dunlap Institute is funded by an endowment established by the David Dunlap family and the University of Toronto.
K.D.O. was supported in part by NSF Grant No. 2207267.
T.T.P. acknowledges support from the Extragalactic Astrophysics Research Group at E\"{o}tv\"{o}s Lor\'{a}nd University, funded by the E\"{o}tv\"{o}s Lor\'{a}nd Research Network (ELKH), which was used during the development of this research.
H.A.R. is supported by NSF Partnerships for Research and Education in Physics (PREP) award No. 2216793.
S.M.R. and I.H.S. are CIFAR Fellows.
Portions of this work performed at NRL were supported by ONR 6.1 basic research funding.
J.D.R. also acknowledges support from start-up funds from Texas Tech University.
J.S. is supported by an NSF Astronomy and Astrophysics Postdoctoral Fellowship under award AST-2202388, and acknowledges previous support by the NSF under award 1847938.
C.U. acknowledges support from BGU (Kreitman fellowship), and the Council for Higher Education and Israel Academy of Sciences and Humanities (Excellence fellowship).
A.V. is supported by the UK's Science and Technology Facilities Council, grant number ST/W000946/1 and also acknowledges the support of the Royal Society and Wolfson Foundation.
C.A.W. acknowledges support from CIERA, the Adler Planetarium, and the Brinson Foundation through a CIERA-Adler postdoctoral fellowship.
O.Y. is supported by the National Science Foundation Graduate Research Fellowship under Grant No. DGE-2139292.

Computational resources were provided by University of Birmingham BlueBEAR High Performance Computing facility.

We acknowledge usage of the following software packages:
\texttt{ENTERPRISE} \citep{2019ascl.soft12015E}, \texttt{enterprise extensions} \citep{enterprise}, \texttt{matplotlib} \citep{4160265}, \texttt{numpy} \citep{Harris2020}, \texttt{PINT} \citep{Luo_2021}, \texttt{PTMCMC} \citep{justin_ellis_2017_1037579}, \texttt{Tempo2}
\citep{2012ascl.soft10015H}.

\newpage

\bibliography{refs}
\bibliographystyle{aasjournal}

\appendix

\begin{figure*}
    \centering
    \includegraphics[width=\textwidth]{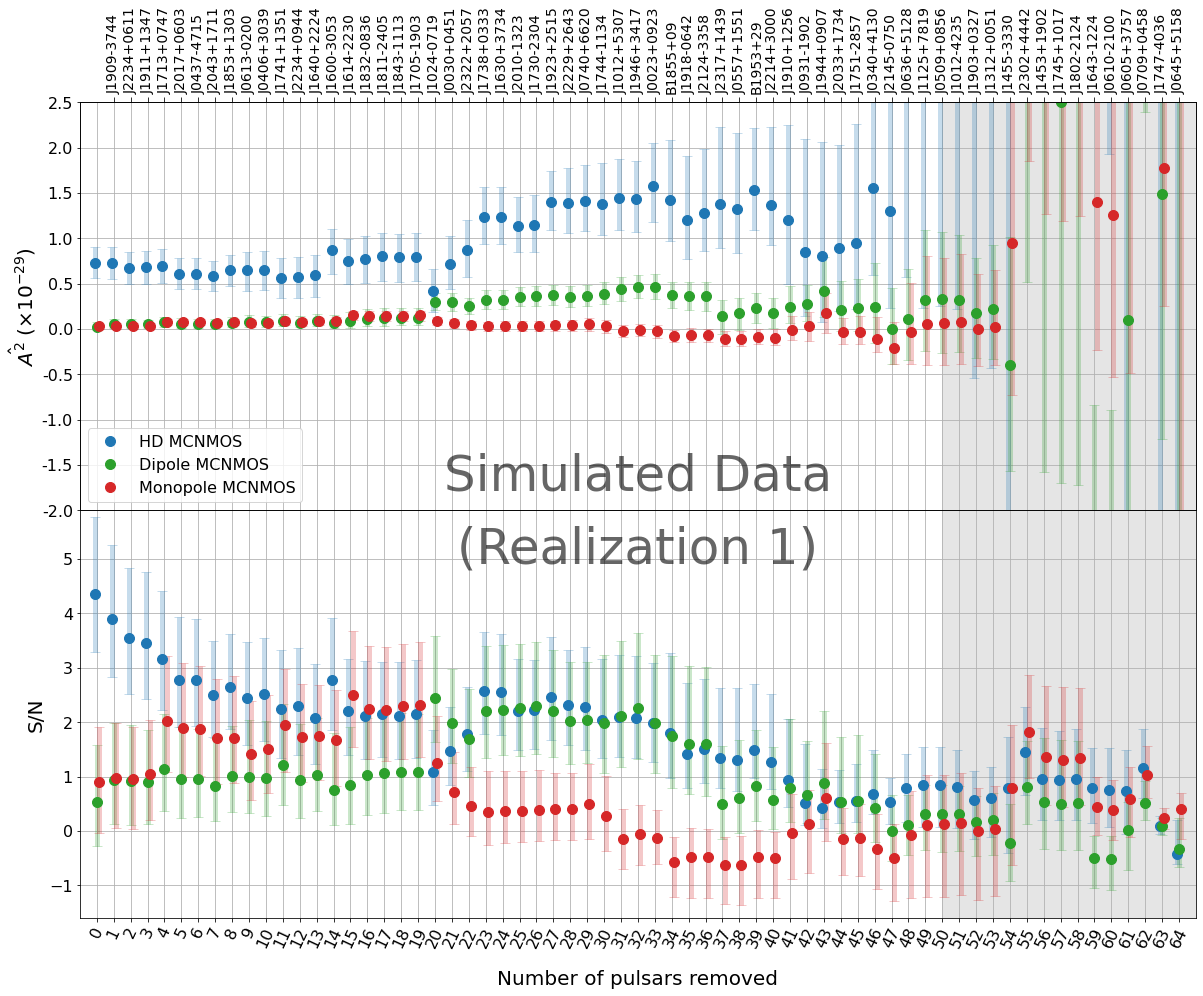}
    \caption{\label{fig:sim_MC_bestfirst}
        The evolution of the MCNMOS and S/N for a simulated NG15 data set (Realization 1), as the least noisy pulsars are removed one by one from the PTA. Similar to Figure~\ref{fig:NG15_SC_bestfirst}; see that caption for more details.
    }
\end{figure*}

\section{Removing the noisiest pulsars first}
\label{app:noisy_first}

\begin{figure*}
    \centering
    \includegraphics[width=\textwidth]{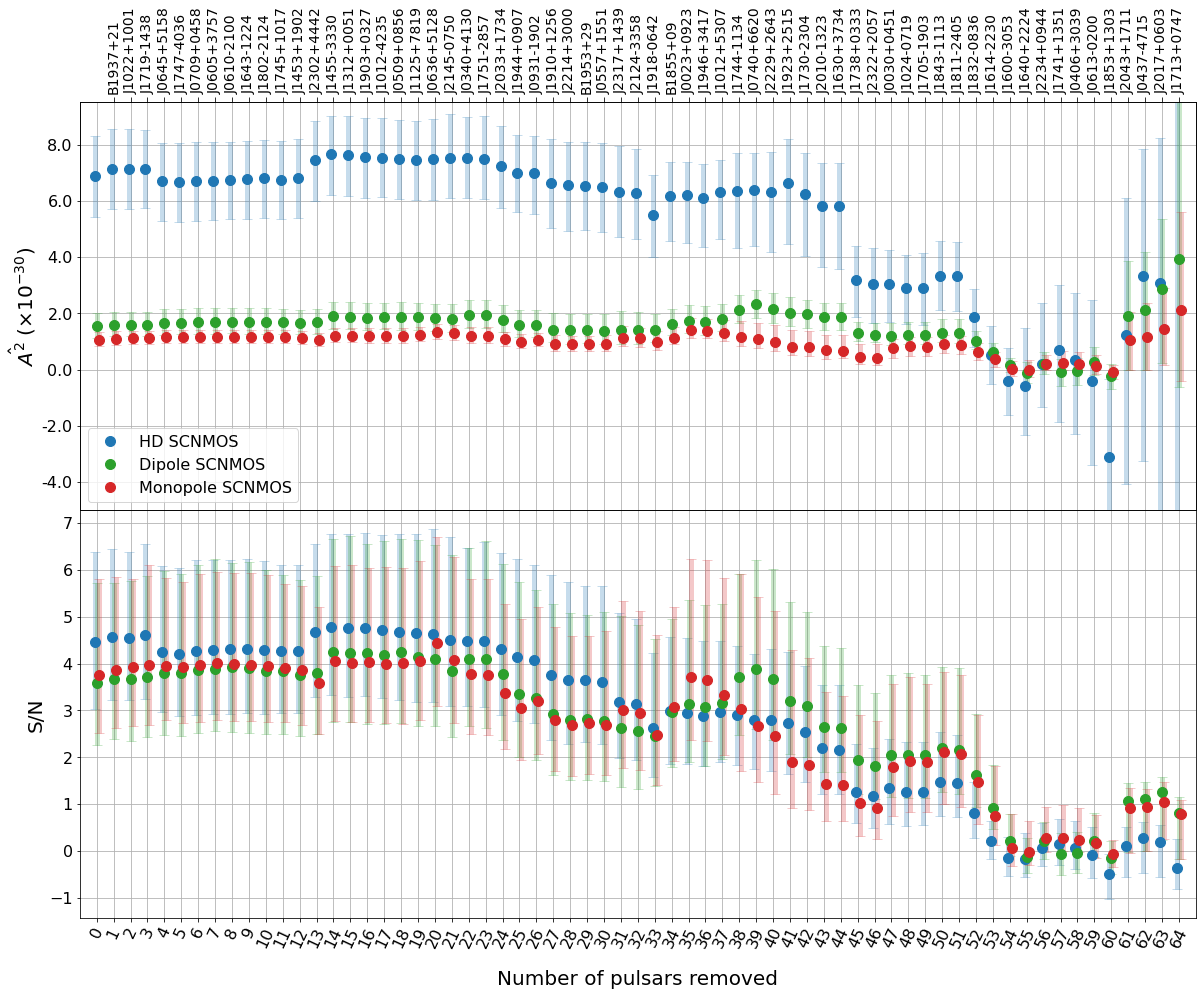}
    \caption{\label{fig:NG15_SC_noisyfirst}
        The evolution of the SCNMOS and S/N for NG15 as the most noisy pulsars are removed one by one from the PTA. Similar to Figure~\ref{fig:NG15_SC_bestfirst}; see that caption for more details.
        In this case we do not show a shaded region beyond 50 pulsars removed (see e.g., Figure~\ref{fig:NG15_SC_bestfirst}). This is because low-noise pulsars remain in the array beyond this point and so the behavior of $\hat{A}^2$ and S/N is less erratic.
    }
\end{figure*}

\begin{figure*}
    \centering
    \includegraphics[width=\textwidth]{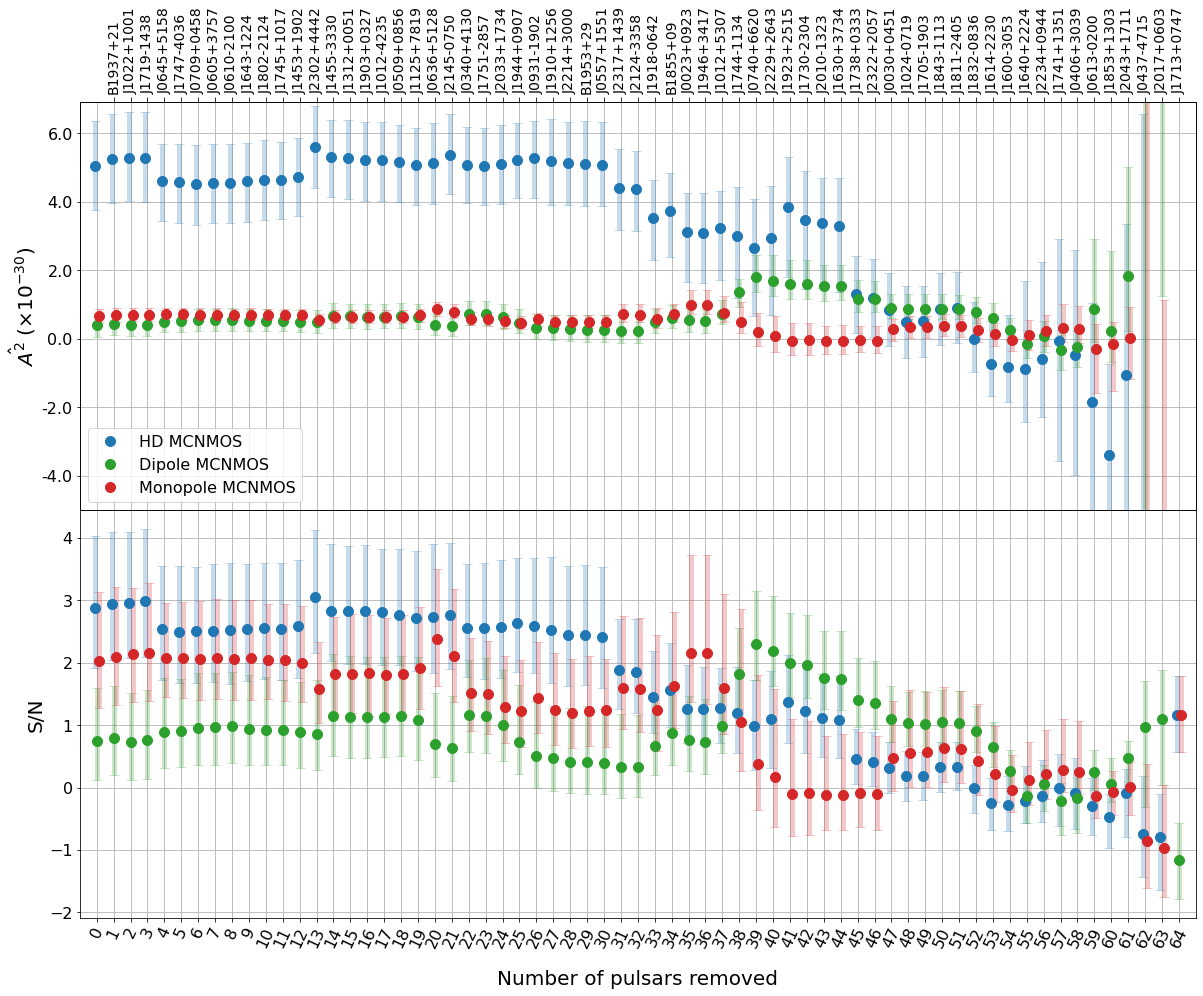}
    \caption{\label{fig:NG15_MC_noisyfirst}
        The evolution of the MCNMOS and S/N for NG15 as the most noisy pulsars are removed one by one from the PTA. Similar to Figure~\ref{fig:NG15_SC_bestfirst}; see that caption for more details. In this case we do not show a shaded region beyond 50 pulsars removed (see e.g., Figure~\ref{fig:NG15_SC_bestfirst}). This is because low-noise pulsars remain in the array beyond this point and so the behavior of $\hat{A}^2$ and S/N is less erratic.
    }
\end{figure*}

In the main body of this work we show analysis results as pulsars are removed from the array, with the least noisy pulsars (those with the smallest white noise TOA uncertainties) removed first. 
This appendix shows the results obtained with the opposite order i.e., when the noisiest pulsars are removed first. Figure~\ref{fig:NG15_SC_noisyfirst} shows how the SCNMOS and associated S/N evolve as pulsars are removed, while Figure~\ref{fig:NG15_MC_noisyfirst} shows the same for the MCNMOS. In both cases $\hat{A}^2$ remains relatively steady until approximately 40 pulsars have been omitted. The S/N is relatively steady until approximately 30 pulsars have been removed.

\section{Distribution of Simulated Features}
\label{app:sim_feat}

\begin{figure*}
   \centering
    \includegraphics[width=0.95\textwidth]{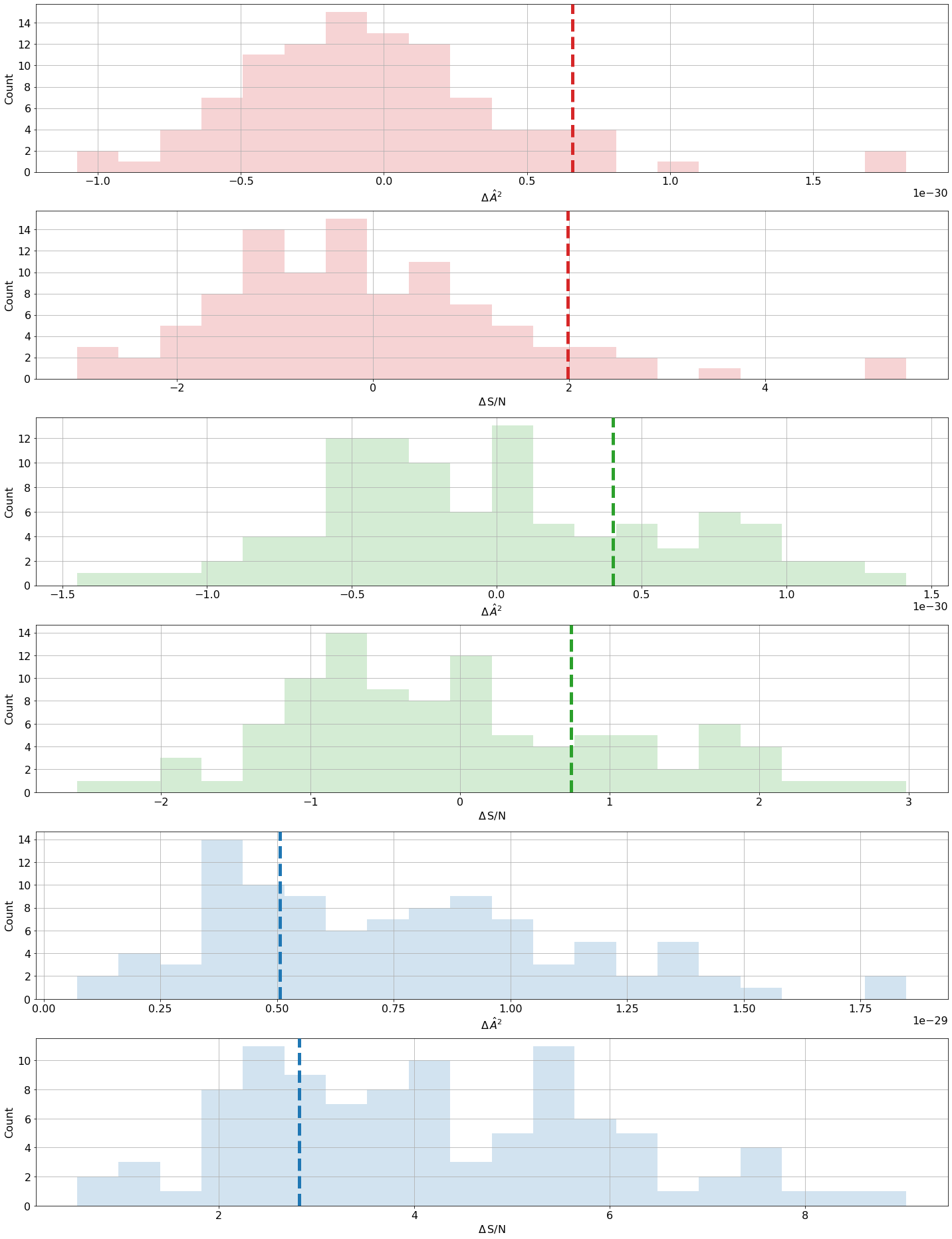}
    \caption{\label{fig:paper_starting}
The distribution of $\hat{A}^2$ and S/N with all 67 NANOGrav pulsars in the array (Metric 1 from Section~\ref{subsec:metrics}). The histograms shows the median values from the MCNMOS analysis for each of the 100 simulated data sets. The results for the monopole, dipole, and HD spatial correlations are red, green and blue respectively. The dashed vertical lines indicate the median values from the MCNMOS analysis of NG15.
    }
\end{figure*}

\begin{figure*}
    \centering
    \includegraphics[width=0.95\textwidth]{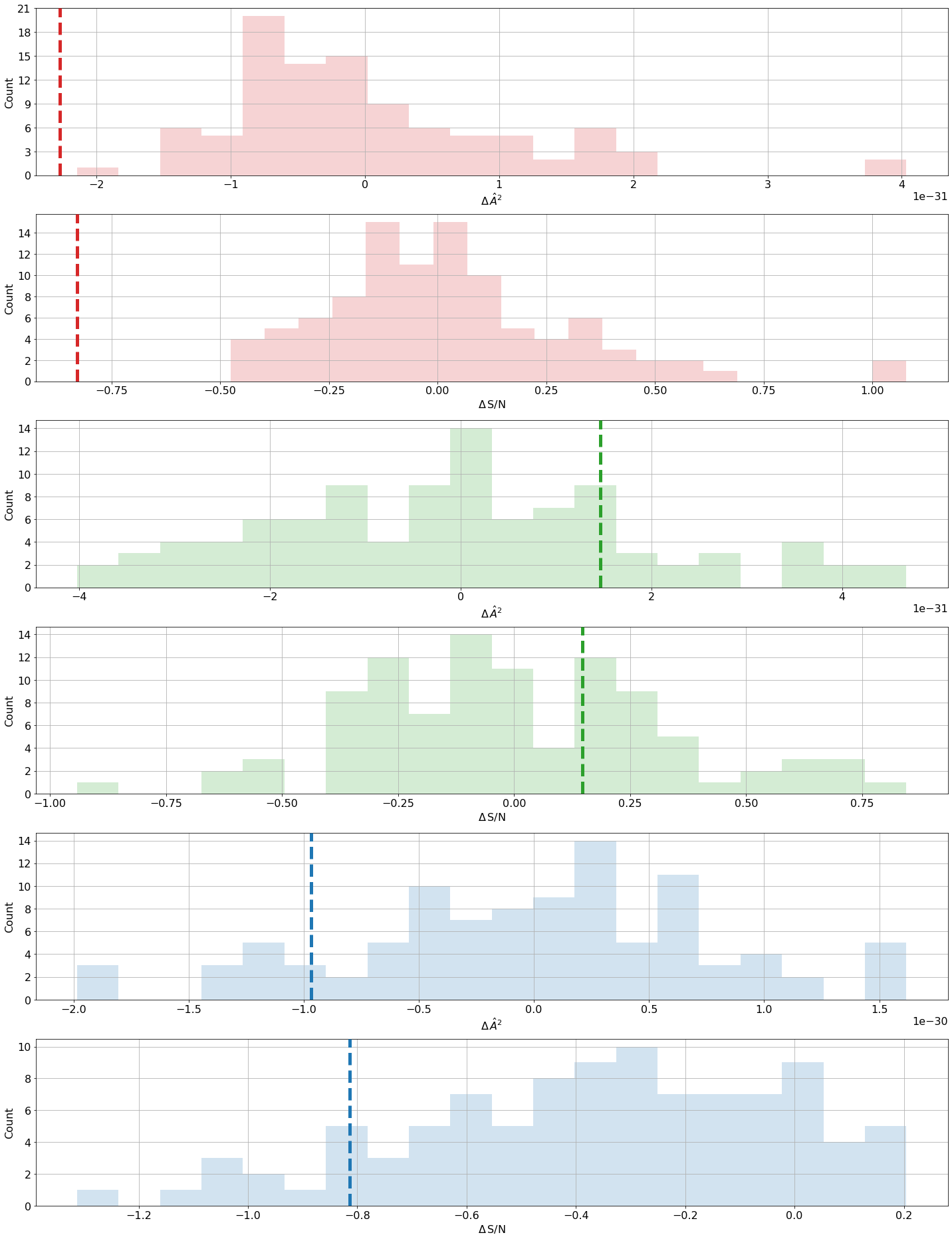}
    \caption{\label{fig:paper_drop}
The distribution of the drop in $\hat{A}^2$ and S/N when the first pulsar is removed (Metric 2 from Section~\ref{subsec:metrics}). Similar to Figure~\ref{fig:paper_starting}; see that caption for more details.
    }
\end{figure*}

\begin{figure*}
    \centering
    \includegraphics[width=0.95\textwidth]{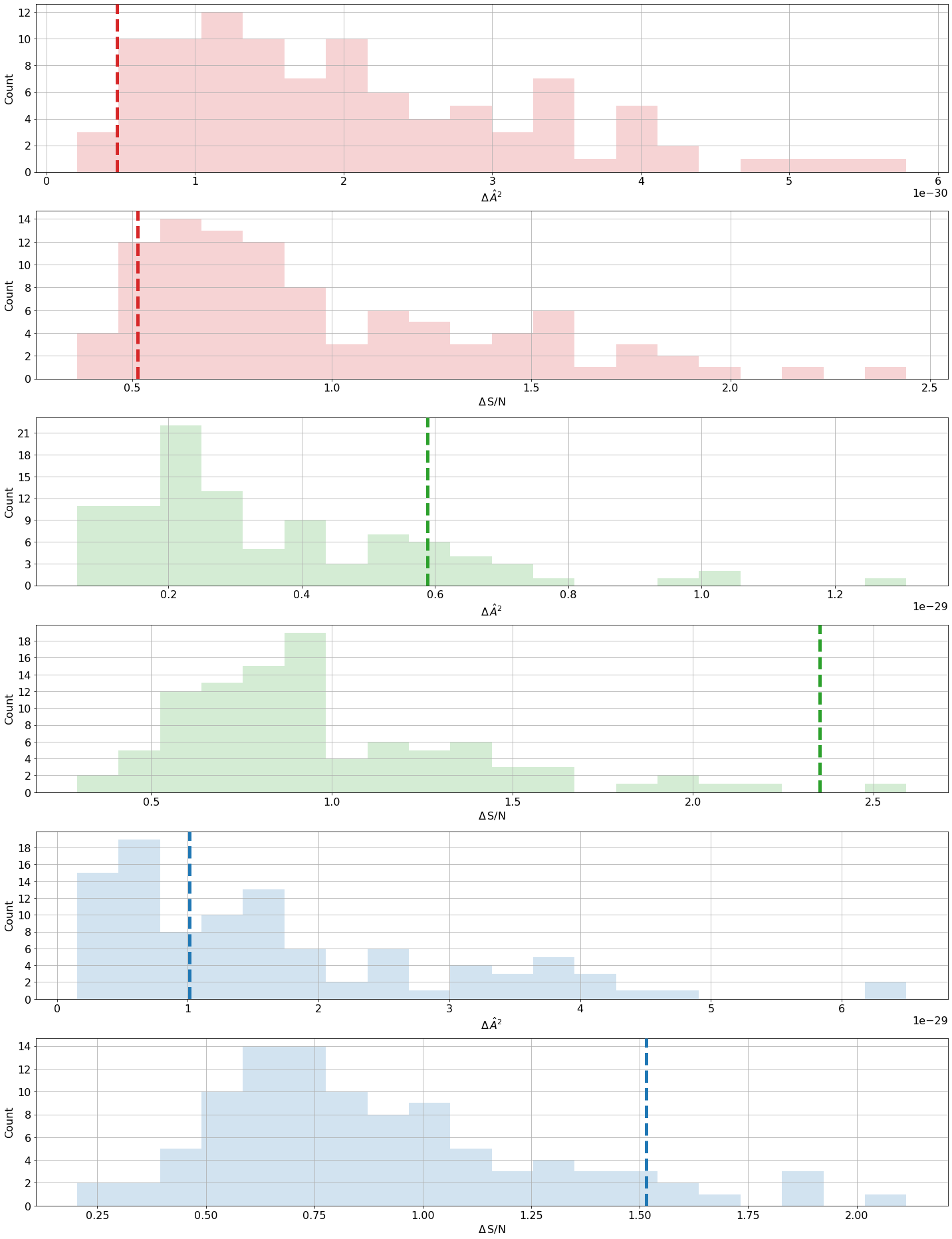}
    \caption{\label{fig:paper_biggest}
The distribution of the largest increase in $\hat{A}^2$ and S/N (Metric 3 from Section~\ref{subsec:metrics}) when a pulsar is removed. Similar to Figure~\ref{fig:paper_starting}; see that caption for more details.
    }
\end{figure*}

Figures~\ref{fig:paper_starting}, \ref{fig:paper_drop} and \ref{fig:paper_biggest} illustrate the distribution in magnitude of various analysis features (as described in Section~\ref{subsec:MC}) across the 100 simulated data sets. The figures also show the magnitude of the corresponding feature for the NG15 data set. Compare with the $p$-values from Table~\ref{tab:p_values}.

\end{document}